%
%
%
%
%
%
%
%
%
%
\hoffset=0.0in
\voffset=0.0in
\hsize=6.5in
\vsize=8.9in
\normalbaselineskip=12pt
\normalbaselines
\topskip=\baselineskip
\parindent=15pt
%
%
%

\let\ep=\epsilon

\let\la=\langle
\let\ra=\rangle

\let\lf=\left
\let\rt=\right
\let\dt=\cdot
\let\del=\nabla

\let\q=\widehat

\let\h=\hbar

\let\rta=\rightarrow

\let\dy=\displaystyle

\let\sy=\scriptstyle
\let\ssy=\scriptscriptstyle
\let\:=\>
\let\\=\cr
\let\emph=\e

\let\m=\hbox

\let\cl=\centerline

\def\e#1{{\it #1\/}}
\def\textbf#1{{\bf #1}}
\def\[{$$}
\def\]{\[}
\def\re#1#2{$$\matrix{#1\cr}\eqno{({\rm #2})}$$}
\def\de#1{$$\matrix{#1\cr}$$}

\def\eqdf{\buildrel{\rm def}\over =}
\def\hf{{\sy {1 \over 2}}}

\def\hfs{{\ssy {1 \over 2}}}

\def\qV{\q{V}}
\def\qcV{\q{{\cal V}}}

\def\qH{\q{H}}

\def\mathrm#1{{\rm #1}}

\def\mathcal#1{{\cal #1}}

\def\mbf{\fam\bffam\tenbf}
\def\bv#1{{\mbf #1}}

\def\vr{\bv{r}}

\def\vp{\bv{p}}

\def\qvp{\q{\vp}}

\def\Schr{Schr\"{o}\-ding\-er}
\font\frtbf = cmbx12 scaled \magstep1
\font\twlbf = cmbx12
\font\ninbf = cmbx9
\font\svtrm = cmr17
\font\twlrm = cmr12
\font\ninrm = cmr9
\font\ghtrm = cmr8

\def\gr#1{{\ghtrm #1}}

\def\abstract#1{{\ninbf\cl{Abstract}}\medskip
\openup -0.1\baselineskip
{\ninrm\leftskip=2pc\rightskip=2pc\noindent #1\par}
\normalbaselines}
\def\sct#1{\vskip 1.33\baselineskip\noindent{\twlbf #1}\medskip}

\def\so{\raise 0.65ex \m{\sevenrm 1}}
\def\sk{\par\vskip 0.66\baselineskip}
{\svtrm
\cl{Nonperturbational ``Continued-Fraction'' Spin-offs of}
\medskip
\cl{Quantum Theory's Standard Perturbation Methods}
}
\bigskip
{\twlrm
\cl{Steven Kenneth Kauffmann\footnote{${}^\ast$}{\gr{Retired, American
Physical Society Senior Life Member, E-mail: SKKauffmann@gmail.com}}}
}
\bigskip\smallskip
\abstract{%
The inherently homogeneous stationary-state and time-dependent
Schr\"{o}dinger equations are often recast into inhomogeneous form in
order to resolve their solution nonuniqueness.  The inhomogeneous term
can impose an initial condition or, for scattering, the preferred per%
mitted asymptotic behavior.  For bound states it provides sufficient
focus to exclude all but one of the homogeneous version's solutions.
Because of their unique solutions, such inhomogeneous versions of
Schr\"{o}dinger equations have long been the indispensable basis for a
solution scheme of successive perturbational corrections which are an%
chored by their inhomogeneous term.  Here it is noted that every such
perturbational solution scheme for an inhomogeneous linear vector
equation spins off a nonperturbational continued-fraction scheme.
Unlike its representation-independent antecedent, the spin-off scheme
only works in representations where all components of the equation's
inhomogeneous term are nonzero.  But that requirement seems to confer
theoretical physics robustness heretofore unknown: for quantum fields
the order of the perturbation places a bound on unperturbed particle
number, the spin-off scheme contrariwise has only basis elements of
unbounded unperturbed particle number.  It furthermore is difficult to
visualize such a continued-fraction spin-off scheme generating infini%
ties, since its successive iterations always go into denominators.
}

\sct{Introduction}
\noindent
\Schr\ equations, whether stationary-state or time-dependent, are ho%
mogeneous, and as such can suffer from solution nonuniqueness.  In or%
der to be able to effectively apply standard successive approximation
schemes, such as perturbational ones, to \Schr\ equations, they are
often first recast into a specialized \e{inhomogeneous} linear form
which has a \e{unique} solution.  Denoting its state vector as $|\psi
\ra$, a simple generic presentation of a \Schr\ equation in inhomogen%
eous form is,
\re{
    |\psi\ra = |\psi_0\ra - \qcV|\psi\ra.
}{1a}
The state vector $|\psi_0\ra$, which comprises the \e{inhomogeneous
term} of Eq.~(1a) usually shares a \e{key feature} with $|\psi\ra$,
such as asymptotic behavior in a region of infinite extent or exact
value at an initial time, or, alternately, \e{is a passable approxi%
mation to} $|\psi\ra$.  Given that the Hamiltonian operator of the ho%
mogeneous \Schr\ equation which underlies Eq.~(1a) is $\qH$, we \e{in
addition} assume the existence of an \e{exactly diagonalizable Hamil%
tonian operator} $\qH_0$ that has $|\psi_0\ra$ as one of its \e{eigen%
states}, and for which the linear operator $\qcV$ of Eq.~(1a) is of no
less than \e{first order in} $\qV\eqdf(\qH - \qH_0)$.

If we now take the viewpoint that the inhomogeneous term $|\psi_0\ra$
on the right-hand side of Eq.~(1a) is a sufficiently good approxima%
tion to $|\psi\ra$ on its left-hand side that the remainder term
$-\qcV|\psi\ra$ \e{is only a small perturbation}, then the \e{pertur%
bational successive approximations},
\re{
    |\psi^{(n + 1)}\ra = |\psi_0\ra - \qcV|\psi^{(n)}\ra,
}{1b}
where, of course, $|\psi^{(0)}\ra = |\psi_0\ra$, would be expected to
converge reasonably rapidly.

Additional insight into the Eq.~(1b) \e{perturbational iteration
scheme} can be obtained from taking the second term on the right-hand
side of Eq.~(1a) to its left-hand side, which yields,
\re{
    |\psi\ra + \qcV|\psi\ra = |\psi_0\ra.
}{1c}
Eq.~(1c) has the formal operator solution,
\re{
    |\psi\ra = [1 + \qcV]^{-1}|\psi_0\ra.
}{1d}
This presents the issue of how to practically evaluate the formal op%
erator $[1 + \qcV]^{-1}$.  One way is to try to use its formal geomet%
ric series expansion,
\re{
    [1 + \qcV]^{-1} = 1 - \qcV + (\qcV)^2 +\cdots+ (-\qcV)^n +\cdots,
}{1e}
which is readily verified to yield the exactly the same result when
inserted into Eq.~(1d) as is obtained from successively applying the
Eq.~(1b) perturbational iteration scheme.  Given the well-known con%
vergence and divergence characteristics of the geometric series, it is
apparent that the Eq.~(1b) perturbational iteration scheme for the in%
homogeneous Eq.~(1a) \e{is unlikely to be greatly useful unless}
$|\psi_0\ra$ \e{is substantially dominant over} $\qcV|\psi_0\ra$.

That naturally raises the question of whether there might be an \e{al%
ternate} general iteration approach to the inhomogeneous Eq.~(1a) that
could come to the rescue when its Eq.~(1b) \e{perturbational} itera%
tions converge too slowly or diverge.  A Lippmann-Schwinger nonrela%
tivistic potential-scattering variant of the inhomogeneous Eq.~(1a),
considered for \e{arbitrarily strong} potentials in a recent article~%
[1], has revealed that \e{if} a complete representation basis set $\{
\la\rho_i|\}$ can be found such that \e{every component} $\la\rho_i|
\psi_0\ra$ of the inhomogeneous-term state vector $|\psi_0\ra$ \e{is
nonzero}, then a robust nonperturbational iteration scheme can be de%
vised from \e{the explicit presentation of} Eq.~(1c) \e{in that repre%
sentation}, namely from the specific equalities,
\re{
    \la\rho_i|\psi\ra + \la\rho_i|\qcV|\psi\ra = \la\rho_i|\psi_0\ra.
}{2a}
If we now make the \e{assumption} that $\la\rho_i|\psi\ra$ is, like
$\la\rho_i|\psi_0\ra$, \e{nonzero}, we can \e{factor the left-hand
side of} Eq.~(2a) into $\la\rho_i|\psi\ra$ and \e{the resulting non%
zero cofactor}.  We then proceed to \e{divide both sides of} Eq.~(2a)
\e{by that nonzero cofactor}, which changes the appearance of Eq.~(2a)
to,
\re{
    \la\rho_i|\psi\ra = \la\rho_i|\psi_0\ra/\lf[1 +
    \lf(\la\rho_i|\psi\ra\rt)^{-1}\la\rho_i|\qcV|\psi\ra\rt],
}{2b}
from which we straightforwardly devise the manifestly nonperturbation%
al successive-approximation scheme,
\re{
    \la\rho_i|\psi^{(n + 1)}\ra = \la\rho_i|\psi_0\ra/\lf[1 +
    \lf(\la\rho_i|\psi^{(n)}\ra\rt)^{-1}\la\rho_i|\qcV|\psi^{(n)}\ra\rt],
}{2c}
where, of course, $\la\rho_i|\psi^{(0)}\ra = \la\rho_i|\psi_0\ra$.  If
$\la\rho_i|\psi^{(n)}\ra$ is nonzero, then because $\la\rho_i|\psi_0
\ra$ is nonzero, barring the pathological occurrence of a divergence
in $\la\rho_i|\qcV|\psi^{(n)}\ra$, $\la\rho_i|\psi^{(n + 1)}\ra$ will
in turn be nonzero.  We also note that the iteration scheme of Eq.~%
(2c) has the desirable nonperturbational character of a \e{continued
fraction}.

Thus the Eq.~(1a) generic specialized \e{inhomogeneous} form of the
\Schr\ equation \e{always} spawns \e{not only} the perturbational suc%
cessive-approximation scheme of Eq.~(1b), but \e{as well} the nonper%
turbational continued-fraction successive-approximation scheme of Eq.%
~(2c) that can regarded as its spin-off.

We shall now survey some of the well-known circumstances where a
\Schr\ equation with Hamiltonian operator $\qH$ for a state $|\psi\ra$
is combined \e{both} with a \e{state} $|\psi_0\ra$ which has a crucial
similarity to $|\psi\ra$ \e{and as well} with an exactly diagonaliz%
able \e{Hamiltonian operator} $\qH_0$ that includes $|\psi_0\ra$ as
one of its eigenstates to produce an \e{inhomogeneous} linear equation
for $|\psi\ra$ \e{which has the generic form given by} Eq.~(1a), where
the linear operator $\qcV$ is of at least first order in $\qV\eqdf(\qH
- \qH_0)$.  We shall as well mention some of the complete \e{represen%
tation basis sets} $\{\la\rho_i|\}$ for which certain of the surveyed
inhomogeneous terms $|\psi_0\ra$ have \e{exclusively nonzero compon%
ents} $\la\rho_i|\psi_0\ra$---note that the \e{eigenstate set} of the
Hamiltonian operator $\qH_0$ is \e{entirely unacceptable} as such a
complete representation basis set $\{\la\rho_i|\}$ because \e{all} of
its members \e{aside from} $\la\psi_0|$ \e{itself} are \e{orthogonal}
to $|\psi_0\ra$.

\sct{The Lippmann-Schwinger equation for nonrelativistic potential
scattering}
\noindent
As in Ref.~[1] we consider the coordinate-representation nonrelativis%
tic \Schr\ equation for an eigenstate $\la\vr|\psi_E\ra$ of \e{posi%
tive energy} $E$,
\re{
 \lf(-\h^2\del_\vr^2/(2m) + V(\vr)\rt)\la\vr|\psi_E\ra = E\la\vr|\psi_E\ra,
}{3a}
where,
\re{
    {\dy\lim_{|\vr|\rta\infty}V(\vr) = 0.}
}{3b}
Thus for sufficiently large $|\vr|$, the \Schr\ equation of Eq.~(3a)
reduces to,
\re{
    \lf(-\h^2\del_\vr^2/(2m)\rt)\la\vr|\psi_E\ra = E\la\vr|\psi_E\ra,
}{3c}
which is satisfied by \e{any plane wave} $e^{i\vp\dt\vr/\h}$ for which
$|\vp| = (2mE)^\hfs$ and by \e{any linear superposition} of these plane
waves.  Among those linear superpositions are all the angularly modu%
lated ingoing and outgoing \e{spherical waves} that have wave number
$k = (2mE)^\hfs/\h$.

Now a \e{scattering experiment} at energy $E > 0$ is described by a
\e{particular solution} of the \Schr\ Eq.~(3a) which at sufficiently
large $|\vr|$ that Eq.~(3a) is well-approximated by Eq.~(3c) consists
of only a \e{single} specified plane wave $e^{i\vp\dt\vr/\h}$ of mo%
mentum $\vp$ plus \e{only outgoing} spherical waves~[2].  We denote
this \e{scattering solution} of the \Schr\ Eq.~(3a) as $\la\vr|
\psi^+_\vp\ra$.

It turns out that an \e{inhomogeneous modification} of the \Schr\ Eq.%
~(3a) describes $\la\vr|\psi^+_\vp\ra$ \e{uniquely}, namely the fol%
lowing nonrelativistic Lippmann-Schwinger equation for potential scat%
tering~[2],
\re{
    \la\vr|\psi^+_\vp\ra = e^{i\vp\dt\vr/\h} -
    \la\vr|(\qH_0 - E_\vp - i\ep)^{-1}\qV|\psi^+_\vp\ra,
}{3d}
where $\qH_0\eqdf |\qvp|^2/(2m) = -\h^2\q{\del^2}/(2m)$ is the kinetic
energy \e{operator} and $E_\vp\eqdf |\vp|^2/(2m)$ is the kinetic ener%
gy \e{c-number scalar} that corresponds to the c-number momentum vec%
tor $\vp$.

Taking $\la\vr|\vp\ra\eqdf e^{i\vp\dt\vr/\h}$, we can write Eq.~(3d)
in the form of Eq.~(1a), i.e.,
\re{
    |\psi^+_\vp\ra = |\vp\ra -
    (\qH_0 - E_\vp - i\ep)^{-1}\qV|\psi^+_\vp\ra,
}{3e}
where $|\psi_0\ra = |\vp\ra$ and $\qcV = (\qH_0 - E_\vp - i\ep)^{-1}
\qV$.  Furthermore, since $\qH_0|\vp\ra = E_\vp|\vp\ra$, we can recov%
er the \Schr\ Eq.~(3a) from Eq.~(3e) by multiplying the latter through
by $(\qH_0 - E_\vp)$, followed by rearrangement of the resulting terms
between the left-hand and right-hand sides.

The negative imaginary infinitesimal $-i\ep$ that appears in the Lipp%
mann-Schwinger Eqs.~(3d) and (3e) ensures that \e{only outgoing}
spherical waves are present for sufficiently large $|\vr|$, in \e{ad%
dition}, of course, to the \e{single} specified plane wave $e^{i\vp\dt
\vr/\h}$ of momentum $\vp$.

The usual \e{perturbational iteration} of the Lippmann-Schwinger Eq.~%
(3d) is,
\re{
    \la\vr|\psi^{(n + 1)+}_\vp\ra = e^{i\vp\dt\vr/\h} -
    \la\vr|(\qH_0 - E_\vp - i\ep)^{-1}\qV|\psi^{(n)+}_\vp\ra,
}{3f}
which with $\la\vr|\psi^{(0)+}_\vp\ra = e^{i\vp\dt\vr/\h}$ generates
the familiar perturbational geometric \e{Born series} for nonrelativ%
istic potential scattering~[3].

For \e{nonperturbational} ``continued-fraction'' iteration of the
Lippmann-Schwinger Eq.~(3d) we happen to be in the extraordinarily
fortunate situation that \e{coordinate representation} of the inhomo%
geneous term $|\psi_0\ra =|\vp\ra$ is \e{always nonzero} because $\la
\vr|\vp\ra = e^{i\vp\dt\vr/\h}\ne 0$.  Therefore in the manner of Eq.%
~(1c) and Eqs.~(2a)--(2c), the Lippmann-Schwinger Eq.~(3d) in coordin%
ate representation gives rise to the nonperturbational ``continued
fraction'' iteration scheme,
\re{
    \la\vr|\psi^{(n +1)+}_\vp\ra = e^{i\vp\dt\vr/\h}/\lf[1 +
    \lf(\la\vr|\psi^{(n)+}_\vp\ra\rt)^{-1}
    \la\vr|(\qH_0 - E_\vp - i\ep)^{-1}\qV|\psi^{(n)+}_\vp\ra\rt],
}{3g}
where we of course have that $\la\vr|\psi^{(0)+}_\vp\ra = e^{i\vp\dt
\vr/\h}$.

\sct{A bound-state inhomogeneous equation suitable for iteration solu%
tion}
\noindent
The just-discussed Lippmann-Schwinger equation is an \e{inhomogeneous}
\Schr -equation variant that is of the form of our generic Eq.~(1a).
With an arbitrary plane wave as its \e{inhomogeneous term}; the Lipp%
mann-Schwinger equation enables \e{completely prescribed iteration re%
finement} of that approximating plane wave toward its \e{unique} asso%
ciated ``asymptotically outgoing-only spherical-wave'' \e{exact} scat%
tering wave function.  The \e{key ingredient} that is needed to fash%
ion the \e{inhomogeneous} Lippmann-Schwinger equation from its under%
lying \e{homogeneous} \Schr\ equation is of course the ``outgoing-only
spherical-wave'' free-particle propagator, which is constructed from
the \e{selfsame} free-particle Hamiltonian for which that approximat%
ing plane wave \e{is an eigenstate}.

In contrast, the \e{standard} \Schr\ approach to the perturbational
refinement of an approximating \e{bound state} $|\psi^0_j\ra$ fails to
\e{explicitly present} the \e{inhomogeneous} \Schr -equation variant
which \e{links} that approximating bound state to a \e{unique} associ%
ated \e{exact} bound-state solution $|\psi_j\ra$ of the underlying ho%
mogeneous stationary-state \Schr\ equation $\qH|\psi_j\ra = E_j|\psi_j
\ra$.  Due to the \e{absence} of the information which is \e{inherent}
to that \e{inhomogeneous linking equation}, the standard \Schr\ bound%
-state perturbational approach \e{can't offer minutely prescribed it%
eration instructions} for its perturbational refinement procedure~[4].
More importantly, if that inhomogeneous linking equation is \e{not} in
hand, there is no obvious way to devise alternate \e{nonperturbation%
al} methods for the iteration refinement of approximating bound
states.

Fortunately, we can follow the blueprint that is provided to us by the
Lippmann-Schwinger equation to devise its missing bound-state analog.
A \e{major difference} between the two, however, is that bound states
certainly \e{don't} feature outgoing, ingoing or \e{any other type of
traveling wave}.  Therefore the kind of propagator we need for bound
states is of the \e{standing-wave} type.  Nor can a propagator suited
to an approximating \e{bound state} possibly be constructed from the
\e{free-particle} Hamiltonian operator as it \e{properly is} for the
approximating \e{plane waves} of the Lippmann-Schwinger equation.  The
construction of the \e{standing wave} propagator we need must be from
a Hamiltonian operator $\qH_0$ which has that \e{approximating bound
state} $|\psi^0_j\ra$ as one of its \e{bound-state eigenstates}, i.e.,
$\qH_0$ must satisfy,
\re{
    \qH_0|\psi^0_j\ra = E^0_j|\psi^0_j\ra.
}{4a}
Of course, since that approximating bound state $|\psi^0_j\ra$ is in%
deed \e{bound}, we can \e{normalize it to unity}, i.e.,
\re{
    \la\psi^0_j|\psi^0_j\ra = 1.
}{4b}
Finally, the \e{construction} of the required standing-wave propagator
from the Hamiltonian operator $\qH_0$ \e{can't be accomplished in
practice unless} $\qH_0$ \e{is exactly diagonalizable}.  If all the
eigenstates and eigenvalues of $\qH_0$ \e{are actually available}, its
corresponding standing-wave propagator can be obtained in the schemat%
ic form,
\re{
    {\dy \lf[P/(\qH_0 - E^0_j)\rt]\eqdf
    \lim_{\ep\rta 0}\lf(\qH_0 - E^0_j\rt)
    \lf[\lf(\qH_0 - E^0_j\rt)^2 + \ep^2\rt]^{-1} =
    \sum_{i\ne j}|\psi^0_i\ra\la\psi^0_i|/(E^0_i - E^0_j)},
}{4c}
where the letter $P$ in the standing-wave propagator's definition de%
notes ``principal value''.  The exactly diagonalizable Hamiltonian op%
erator $\qH_0$ is frequently called the ``unperturbed'' Hamiltonian
operator, its particular bound eigenstate $|\psi^0_j\ra$ the ``unper%
turbed'' approximating bound state, and that state's $\qH_0$ eigenval%
ue $E^0_j$ the corresponding ``unperturbed'' approximating bound-state
energy.  By the same token, one could call the operator $\lf[P/(\qH_0
- E^0_j)\rt]$ the ``unperturbed'' standing-wave propagator.  Note that
$\lf[P/(\qH_0 - E^0_j)\rt]$ has a special feature \e{which has no ana%
log for traveling-wave propagators} such as the outgoing-wave propaga%
tor of the Lippmann-Schwinger equation, namely that,
\re{
    \la\psi^0_j|\lf[P/(\qH_0 - E^0_j)\rt] = 0.
}{4d}
The final ingredient we need to obtain the bound-state analog of the
inhomogeneous Lippmann-Schwinger equation is the homogeneous station%
ary-state \Schr\ equation for the \e{exact} bound state $|\psi_j\ra$
that is \e{uniquely linked} to the ``unperturbed'' approximating
bound state $|\psi^0_j\ra$.  Denoting the \e{exact} Hamiltonian oper%
ator for the physical system we are studying as $\qH$, this homogen%
eous stationary-state \Schr\ equation is of course,
\de{\qH|\psi_j\ra = E_j|\psi_j\ra,}
which it can sometimes be more convenient to express as,
\de{\lf[\qH - E_j\rt]|\psi_j\ra = 0.}
Since the physical system's \e{exact} Hamiltonian operator $\qH$
\e{won't} in general be exactly diagonalizable (\e{unlike} the some%
what artificial ``unperturbed'' Hamiltonian operator $\qH_0$), we have
no more a priori knowledge of the particular \e{exact} bound-state
\e{energy} $E_j$ than we do of its corresponding \e{exact} bound-state
eigenstate $|\psi_j\ra$ of the exact Hamiltonian operator $\qH$.  In
order to \e{avoid} having the exact bound-state energy $E_j$ become a
\e{completely independent unknown for which we need to solve}, we
choose to write the physical system's exact homogeneous stationary-%
state \Schr\ equation for the bound state $|\psi_j\ra$ in the more
cumbersome form,
\re{
    \lf[\qH - \lf(\la\psi_j|\qH|\psi_j\ra/
     \la\psi_j|\psi_j\ra\rt)\rt]|\psi_j\ra = 0,
}{4e}
which \e{entirely supplants} $E_j$ by $|\psi_j\ra$ and $\qH$.

Eqs.~(4a)--(4e) supply all the ingredients needed to devise the bound%
-state analog of the Lippmann-Schwinger equation.  Inspection of the
Lippmann-Schwinger equation strongly suggests that we multiply the
physical system's homogeneous \Schr\ Eq.~(4e) from the left by the
``unperturbed'' standing-wave propagator $\lf[P/(\qH_0 - E^0_j)\rt]$
as the first step toward that analog.  That operator multiplication
produces the physical \Schr -equation variant,
\re{
    \lf[P/(\qH_0 - E^0_j)\rt]\lf[\qH - \lf(\la\psi_j|\qH|\psi_j\ra/
     \la\psi_j|\psi_j\ra\rt)\rt]|\psi_j\ra = 0,
}{4f}
which is not yet the \e{inhomogeneous} Lippmann-Schwinger equation an%
alog that we seek.  To complete our task we now develop an \e{identity}
involving $|\psi_j\ra$, $\lf[P/(\qH_0 - E^0_j)\rt]$ and $\lf[-\qH_0 +
E^0_j\rt]$ to which we shall then \e{add} the Eq.~(4f) physical
\Schr -equation variant.  One might naively expect that,
\de{|\psi_j\ra + \lf[P/(\qH_0 - E^0_j)\rt]\lf[-\qH_0 + E^0_j\rt]|\psi_j\ra =
0,}
but if we multiply the left-hand side of this proposed equality by
$\la\psi^0_j|$ and take note of Eq.~(4d), we see that the result is
$\la\psi^0_j|\psi_j\ra$ rather than zero.  From Eqs.~(4c) and (4b) it is
indeed seen that,
\re{
    |\psi_j\ra + \lf[P/(\qH_0 - E^0_j)\rt]\lf[-\qH_0 + E^0_j\rt]|\psi_j\ra =
    |\psi^0_j\ra\la\psi^0_j|\psi_j\ra.
}{4g}
We now add the Eq.~(4g) \e{identity} to the physical \Schr -equation
variant given by Eq.~(4f) to obtain,
\re{
    |\psi_j\ra +
    \lf[P/(\qH_0 - E^0_j)\rt]\lf[\qH - \qH_0 + E^0_j
    - \lf(\la\psi_j|\qH|\psi_j\ra/
     \la\psi_j|\psi_j\ra\rt)\rt]|\psi_j\ra =
    |\psi^0_j\ra\la\psi^0_j|\psi_j\ra.
}{4h}
If we now define the ``interaction Hamiltonian operator'' $\qV$ as,
\re{
    \qV\eqdf (\qH - \qH_0),
}{4i}
we can rewrite Eq.~(4h) as,
\re{
 |\psi_j\ra +
 \lf[P/(\qH_0 - E^0_j)\rt]\lf[\qV -
 \lf(\la\psi_j|(\qV + \qH_0 - E^0_j)|\psi_j\ra/
 \la\psi_j|\psi_j\ra\rt)\rt]|\psi_j\ra =
 |\psi^0_j\ra\la\psi^0_j|\psi_j\ra.
}{4j}
If we move the term involving the ``unperturbed'' standing-wave propa%
gator $\lf[P/(\qH_0 - E^0_j)\rt]$ to the right-hand side of Eq.~(4j),
we obtain,
\re{
 |\psi_j\ra =
 |\psi^0_j\ra\la\psi^0_j|\psi_j\ra -
 \lf[P/(\qH_0 - E^0_j)\rt]\lf[\qV
 - \lf(\la\psi_j|(\qV + \qH_0 - E^0_j)|\psi_j\ra/
 \la\psi_j|\psi_j\ra\rt)\rt]|\psi_j\ra,
}{4k}
which is the inhomogeneous bound-state analog of the Lippmann-Schwing%
er equation presented in Eq.~(3e).  The exact energy $E_j$ of the exact
physical bound state $|\psi_j\ra$ is of course given by,
\re{
 E_j = \lf(\la\psi_j|(\qV + \qH_0)|\psi_j\ra/
 \la\psi_j|\psi_j\ra\rt).
}{4l}
The complete detailed prescription for the \e{perturbational itera%
tion} of Eq.~(4k) is then obviously,
\re{
 |\psi^{(n + 1)}_j\ra =
 |\psi^0_j\ra\la\psi^0_j|\psi^{(n)}_j\ra -
 \lf[P/(\qH_0 - E^0_j)\rt]\lf[\qV -
 \lf(\la\psi^{(n)}_j|(\qV + \qH_0 - E^0_j)|\psi^{(n)}_j\ra/
 \la\psi^{(n)}_j|\psi^{(n)}_j\ra\rt)\rt]|\psi^{(n)}_j\ra,
}{4m}
where, of course, $|\psi^{(0)}_j\ra = |\psi^0_j\ra$.  In addition,
\e{entirely as a byproduct of} $|\psi^{(n)}_j\ra$, we have,
\re{
 E^{(n + 1)}_j = \lf(\la\psi^{(n)}_j|(\qV + \qH_0)|\psi^{(n)}_j\ra/
 \la\psi^{(n)}_j|\psi^{(n)}_j\ra\rt).
}{4n}

As was pointed out earlier, the complete detailed prescription of
Eqs.~(4m) and (4n) for the perturbational iteration of a bound state
approximation $|\psi^0_j\ra$ cannot be obtained within the confines of
the \e{standard} \Schr\ bound-state perturbation approach~[4] because
that approach makes no attempt to work out the \e{inhomogeneous} Eq.~%
(4j) or (4k) variant of the stationary-state \Schr\ equation for the
exact bound state $|\psi_j\ra$.  Of course having Eq.~(4j) \e{in hand}
is \e{also} absolutely critical for the development of \e{nonpertuba%
tional} ``continued fraction'' iteration of $\la\rho_i|\psi^0_j\ra$,
where $\la\rho_i|\psi^0_j\ra$ must be \e{nonzero} for every member of
the complete orthogonal basis set $\{\la\rho_i|\}$.  The prescription
for that nonperturbational ``continued fraction'' iteration clearly
comes out to be,
\re{
 \la\rho_i|\psi^{(n + 1)}_j\ra =
 \la\rho_i|\psi^0_j\ra\la\psi^0_j|\psi^{(n)}_j\ra/\cr
 \lf\{1 +
 \lf(\la\rho_i|\psi^{(n)}_j\ra\rt)^{-1}
 \la\rho_i|\lf[P/(\qH_0 - E^0_j)\rt]\lf[\qV -
 \lf(\la\psi^{(n)}_j|(\qV + \qH_0 - E^0_j)|\psi^{(n)}_j\ra/
 \la\psi^{(n)}_j|\psi^{(n)}_j\ra\rt)\rt]|\psi^{(n)}_j\ra\rt\},
}{4o}
where, of course, $\la\rho_i|\psi^{(0)}_j\ra = \la\rho_i|\psi^0_j\ra$,
and the approximation $E^{(n + 1)}_j$ to the energy eigenvalue is giv%
en by Eq.~(4n).

For a bound state approximation $|\psi^0_j\ra$, finding a complete or%
thogonal basis set $\{\la\rho_i|\}$ such that every $\la\rho_i|\psi^0
_j\ra$ is \e{nonzero} may not be simple because, aside from the ground
state, bound states normally have \e{nodes} (where they of course van%
ish) in both coordinate and momentum representation.  However \e{if}
that bound state approximation $|\psi^0_j\ra$ is an eigenstate of an
``unperturbed'' \e{harmonic oscillator Hamiltonian} $\qH_0$, then one
can find a related ``harmonic-oscillator coherent false ground-state
with false excitations'' orthogonal basis $\{\la\psi(c)^0_i|\}$ such
that every $\la\psi(c)^0_i|\psi^0_j\ra$ is indeed nonvanishing.  Here
$c$ is a complex number such that $|c|^2$ is a \e{transcendental} pos%
itive real number, e.g., the natural-logarithm base $e$ or the con%
stant $\pi$, and the ``false ground state'' is the minimum-uncertainty
\e{coherent state} that is characterized by the complex number $c$,
i.e., it is annihilated by one of the harmonic-oscillator \e{annihila%
tion operators} with $c$ \e{subtracted from it}.  Successive mutually
orthogonal states $|\psi(c)^0_i\ra$ are then generated from this par%
ticular \e{coherent} ``false ground state'' characterized by $c$
through the repeated action of the \e{corresponding harmonic-oscilla%
tor creation operator} with the \e{complex conjugate} $\bar{c}$ of $c$
subtracted from it---these are the ``false excitations'' of the coher%
ent ``false ground state''.  It turns out that $\la\psi(c)^0_i|
\psi^0_j\ra$ can't vanish if $|c|^2$ is a positive transcendental num%
ber because it only vanishes when $|c|^2$ is a zero of an appropriate
polynomial that has rational coefficients. We note that the \e{non%
transcendental} (i.e., the ``algebraic'') real numbers, being a
\e{countable set}, are of measure zero.  Furthermore, choosing to use
harmonic-oscillator states as bound state approximations is likely a
viable proposition, e.g., for a bound state of a three-dimensional
system, its approximation by a suitable three-dimensional harmonic os%
cillator state could probably be made to work out quite well.

\sct{Time-dependent Dirac-picture successive approximation methods}
\noindent
It is often the case that the Hamiltonian operator $\qH$ for a time-%
dependent \Schr\ equation,
\re{
    i\h d(|\psi(t)\ra)/dt = \qH|\psi(t)\ra,
}{5a}
can be written in the form $\qH = \qH_0 + \qV$, where $\qH_0$ \e{is
exactly diagonalizable} and $|\psi(t_0)\ra$ is specified to equal one
of the eigenstates $|\psi_0\ra$ of $\qH_0$.  In that case it can be
useful to reexpress the time-dependent \Schr\ Eq.~(5a) in the \e{Dirac
picture}~[5],
\re{
    |\psi(t)\ra = e^{-i\qH_0(t - t_0)/\h}|\psi_D(t)\ra,
}{5b}
which yields,
\re{
    i\h d(|\psi_D(t)\ra)/dt = \qV_D(t)|\psi_D(t)\ra, 
}{5c}
where,
\re{
    \qV_D(t)\eqdf e^{+i\qH_0(t -t_0)/\h}\qV e^{-i\qH_0(t - t_0)/\h},
}{5d}
and $|\psi_D(t_0)\ra = |\psi_0\ra$.  Through integration, Eq.~(5c) can
be reexpressed in an \e{inhomogeneous form} that \e{incorporates} the
eigenstate $|\psi_0\ra$ of $\qH_0$ as the \e{inhomogeneous term},
\re{
 |\psi_D(t)\ra = |\psi_0\ra - (i/\h)\int_{t_0}^t\qV_D(t')|\psi_D(t')\ra dt'.
}{5e}
The inhomogeneous Eq.~(5e) has the form of our generic Eq.~(1a), and
therefore is subject to either standard perturbational iteration,
\re{
    |\psi^{(n + 1)}_D(t)\ra = |\psi_0\ra -
    (i/\h)\int_{t_0}^t\qV_D(t')|\psi^{(n)}_D(t')\ra dt',
}{5f}
where of course $|\psi^{(0)}_D(t)\ra = |\psi_0\ra$, or, if we can find
a complete orthogonal basis set $\{\la\rho_i|\}$ such that $\la\rho_i|
\psi_0\ra$ is \e{never zero}, we can carry out \e{nonperturbational}
``continued-fraction'' iteration,
\re{
    \la\rho_i|\psi^{(n + 1)}_D(t)\ra = \la\rho_i|\psi_0\ra/\lf[1 +
    (i/\h)\lf(\la\rho_i|\psi^{(n)}_D(t)\ra\rt)^{-1}
    \int_{t_0}^t\la\rho_i|\qV_D(t')|\psi^{(n)}_D(t')\ra dt'\rt],
}{5g}
where of course $\la\rho_i|\psi^{(0)}_D(t)\ra = \la\rho_i|\psi_0\ra$.

A prime application of the Dirac-picture Eq.~(5e) is to \e{quantum
field theories}~[6], where it has invariably so far been pursued by
using the standard perturbational iteration scheme of Eq.~(5f), no
doubt most famously by Dyson to systematically extract the \e{quintes%
sentially perturbational} Feynman diagrams~[6].  It would certainly
be interesting if the \e{nonperturbational} ``continued-fraction''
iteration scheme of Eq.~(5g) could conceivably be applied to quantum
field theories.

In quantum-field applications $|\psi_0\ra$ is a small number of unper%
turbed boson and fermion particles.  Now unperturbed boson systems can
be regarded as a massive collection of quantized simple harmonic os%
cillators, while unperturbed fermion systems can be regarded as an e%
qually massive collection of quantum two-state entities.  In the prev%
ious section we have pointed out that systems which are initially har%
monic oscillators can be made compatible with continued-fraction iter%
ation by using a ``coherent false ground state'' and related ``false
excitation'' \e{orthogonal basis} (in quantum field theories perhaps
better termed a ``coherent false vacuum'' and related ``false Bose-%
particle'' \e{orthogonal basis}), wherein quantized oscillator annihi%
lation operators are translated by a complex number, and creation op%
erators by that number's complex conjugate, which, of course can
\e{also} be described as a simple \e{real-valued c-number vector
translation of the quantum coordinate-momentum operator phase space}.
The \e{reason} that such c-number operator translations simply gener%
ate \e{additional orthogonal bases} is, of course, that c-number
translations have \e{no effect on the crucial commutator algebra of
those operators}, i.e., they are \e{canonical} transformations, indeed
\e{unitary} ones.

Systems which are initially collections of two-state systems would
seem in principle even easier to make compatible with ``continued-%
fraction'' iteration. In Pauli-matrix language, if the natural basis
for the initial two-state system consists of the two eigenstates of
$\hf(1 + \sigma_z)$ (which has the two eigenvalues $0$ and $1$, namely
the two natural fermionic-state occupation numbers), then for this
particular purpose of ``continued fraction'' iteration, we can \e{in%
stead} adopt the two eigenstates of $\hf(1 + \sigma_y)$, for example,
since each state of the natural basis indeed has a \e{a nonzero inner
product with each one of the states of this adopted basis}.  The exam%
ple just given can be described as a spinor basis rotation of ninety
degrees, and other rotation angles ought to achieve the desired result
as well.  Note that the natural unperturbed-particle basis has here
once again been subjected to \e{a unitary transformation}.

The fact that the continued-fraction iteration scheme \e{forces} on
one a basis that has \e{no zero-valued inner products} with the mem%
bers of the natural particle occupation-number basis of the quantum
field system's unperturbed Hamiltonian operator $\qH_0$ implies that
this iteration scheme \e{will very rapidly indeed} force an \e{un%
bounded number} of unperturbed particles into active participation.
That is in very sharp contrast with the perturbational iteration
scheme: the \e{intrinsic nature} of Feynman diagrams \e{immediately
reveals} that no more than a \e{certain maximum number} of unperturbed
particles \e{can ever participate} through a given order of the per%
turbational iteration.  (The only exception to this iron Feynman-dia%
gram rule occurs when individual diagrams \e{fail by yielding infrared
divergences} -- whose \e{cure} is widely agreed to involve \e{coher%
ent} photon states, \e{which have no upper bound on the possible num%
ber of photons which they describe}).  The continued-fraction itera%
tion scheme, contrariwise, \e{forces on one} coherent states (i.e.,
``false vacua'') and their related mutually orthogonal ``false Bose-%
particle'' brethren that \e{all} have an unbounded number of unper%
turbed particles.  It \e{as well} forces on one the fact that \e{all}
the  fermionic basis states have a greater than zero probability \e{to
actually be occupied by unperturbed fermions}.  Feynman diagrams re%
veal ``virtual particles'', but continued-fraction iteration reveals
immense ``virtual clouds'' \e{which have no upper bound on the number
of unperturbed particles involved}.  Landing in the thick of such
``virtual clouds'' within an iteration or two suggests much better
convergence for the continued-fraction scheme than for the perturba%
tional one.

Feynman diagrams can \e{diverge}.  In the continued-fraction scheme,
everything that is calculated gets pushed into a \e{denominator},
which makes actual divergence quite difficult to visualize mathemati%
cally.  What diverges in the continued-fraction scheme would seem to
be \e{unperturbed-particle participation} rather than the iteration
expressions themselves.

\vskip 1.75\baselineskip\noindent{\frtbf References}
\vskip 0.25\baselineskip

{\parindent = 15pt
\sk\item{[1]}
S. K. Kauffmann,
viXra:1212.0128 Mathematical Physics,
http://vixra.org/pdf/1212.0128v1.pdf
(2012).
\sk\item{[2]}
M. L. Goldberger and K. M. Watson,
\e{Collision Theory}
(John Wiley \& Sons, New York, 1964), pp.~197--199.
\sk\item{[3]}
M. L. Goldberger and K. M. Watson, op.\ cit., pp.~306--313.
\sk\item{[4]}
L. I. Schiff,
\e{Quantum Mechanics}
(McGraw-Hill, New York, 1955), pp.~151--154.
\sk\item{[5]}
L. I. Schiff, op.\ cit., pp.~195--205.
\sk\item{[6]}
S. S. Schweber
\e{An Introduction to Relativistic Quantum Field Theory}
(Harper \& Row, New York, 1961).
}
\bye